\documentclass[10pt]{bmc_article}

\usepackage{cite} 
\usepackage{url}  
\usepackage{ifthen}  
\usepackage{multicol}   
\usepackage[utf8]{inputenc} 
\usepackage{amssymb,amsfonts,amsmath,bbm}
\usepackage[dvips,dvipdfm]{graphicx}
\newboolean{publ}

\newenvironment{bmcformat}{\fussy\setboolean{publ}{true}}{\fussy}

\begin{document}
\begin{bmcformat}


\title{On the role of intrinsic noise on the response of the p53-Mdm2 module}
 

\author{L\'{\i}dice Cruz-Rodr\'{\i}guez$^{1}$%
\email{L\'{\i}dice Cruz-Rodr\'{\i}guez - lcruz@fisica.uh.cu}%
\and
Nuris Figueroa-Morales$^{1}$%
\and
\email{Nuris Figueroa-Morales - nfigueroa@fisica.uh.cu}%
Roberto Mulet$^{1}$%
\email{Roberto Mulet\correspondingauthor - mulet@fisica.uh.cu}%
}

\address{%
\iid(1)``Henri-Poincar\'e-Group'' of Complex Systems and Department of Theoretical Physics, Physics Faculty, University of Havana, La Habana, CP 10400, Cuba
}
\maketitle


\begin{abstract}
\paragraph*{Background:}%

The protein p53 has a well established role in protecting genomic integrity in human cells. When DNA is damaged p53 induces the cell cycle arrest to prevent the transmission of the damage to cell progeny, triggers the production of proteins for DNA repair and ultimately calls for apoptosis.
In particular, the p53-Mdm2 feedback loop seems to be the key circuit in this response of cells to damage. For many years, based on measurements over populations of cells it was believed that the p53-Mdm2 feedback loop was the responsible for the existence of damped oscillations in the levels of p53 and Mdm2 after DNA damage.
However, recent measurements in individual human cells have shown that p53 and its regulator Mdm2 develop sustained oscillations over long periods of time even in the absence of stress. These results have attracted a lot of interest, first because they open a new experimental framework to study the p53 and its interactions and second because they challenge years of mathematical models with new and accurate data on single cells. Inspired by these experiments standard models of the p53-Mdm2 circuit were modified introducing ad-hoc some biologically motivated noise that becomes responsible for the stability of the oscillations. Here, we follow an alternative approach proposing that the noise that stabilizes the fluctuations is the intrinsic noise due to the finite nature of the populations of p53 and Mdm2 in a single cell.

\paragraph*{Results:}%
We study three stochastic models of the p53-Mdm2 circuit. The models  capture the response of the p53-Mdm2 circuit in its basal state, in the presence of DNA damage, and under oncogenic signals. They are studied using Gillespie's simulations, mean field methods and analytical predictions within the context of the Linear Noise Approximation. For the first two models our results compare quantitatively well with existing experimental data in single cells.  The study of the response of the p53-Mdm2 circuit under oncogenic signals is  process for which we do not have single cell measurements, but can  be modeled waiting for experimental verifiable results.  While we can not discard that other sources of noise in the cells may also be important, our results strongly support the relevance of the intrinsic noise in these systems. 

\paragraph*{Conclusions:}%
We suggest that the noise induced by the finite size of the populations is responsible for the existence of sustained oscillations in the response of the p53-Mdm2 circuit. This noise alone can explain most of the experimental results obtained studying the dynamics of the p53-Mdm2 circuit in individual cells.

\end{abstract}

\ifthenelse{\boolean{publ}}{\begin{multicols}{2}}{}



\section*{Background}

The p53 protein has attracted special attention during the last twenty years because there is strong evidence of its malfunction in most human cancers \cite{Bert}. Like most proteins, the p53 may be inactivated directly by mutations in its own gene or as a result of the interactions  with other genes that transmit information to or from the p53.  Unfortunately, the many pathways that control its activation and the comparable large number of functions, some apparently contradictory, still makes the complete understanding of the role played by this protein an unsolved problem \cite{Bert,Melisa}.
 

In general, p53 works as a transcription factor that positively or negatively regulates the expression of several, and very different genes. The p53 gene was first identified as a tumor suppressor gene already many years ago, but we have now evidence that it has a role in the regulation of glycolysis \cite{Karim}, the repair of oncogenic response \cite{Sherr}, the regulation of metabolism  \cite{Karen} and many others\cite{Melisa}. 

It is now well accepted that one of the main mechanisms regulating the expression of the p53 is  its interaction with the protein Mdm2. The p53 and the Mdm2 form a feedback loop very common in many biological systems. In normal conditions the activation of the p53 activates the Mdm2 that in turn suppresses the p53\cite{Michael}. Then, if the cell is in the presence of stress, like DNA damage, p53 and/or Mdm2 are phosphorilated and its interaction is reduced increasing the level of p53 in the cell \cite{Michael}.

An important breakthrough in the study of this system was achieved a few years ago when researchers tracked, cell by cell the concentration of p53 and Mdm2 after irradiation with gamma rays \cite{Zatorsky}. More recent experiments of the same groups were fundamental to clarify  at least the following issues \cite{ Zatorsky,Loewer, Batchelor}:

\begin{itemize}
\item The dynamics of the p53 is dephased from the dynamics of the Mdm2 and both are characterized by a fixed frequency  and variable amplitudes of oscillations within a single cell.
\item The dynamics and the quantity of the protein p53 is similar before and after irradiation. This open the question of how the p53 leads to cycle arrest or apoptosis when the cell is damaged.
\item The levels of p53 are uncoupled from its transcriptional activity.
\end{itemize}

From the mathematical point of view these results inspired the development of new models. Most of them were built on previous ideas about the interactions within the p53-Mdm2 circuit but adding, ad-hoc, some biologically motivated noise that will be responsible for the stability of the oscillations\cite{Zatorsky, Abou,Ouattara}. Others, introduced a delay mechanism between the p53 activation and the appearance of the Mdm2 \cite{Ciliberto,Ma}. 

On the other hand, experiments suggest that the size of the population of p53 in the basal and activated states are rather small, $10^3-10^4$ elements (see for example \cite{Ma}) suggesting than mean field approaches may not be appropriate. Here, we follow an alternative approach and propose that the finite size of the populations involved is the source of an intrinsic noise that, in a single cell, stabilizes the oscillations. This kind of approach was recently suggested in ref. \cite{Liu} but here we push it further. First studying 
not only the isolated p53-Mdm2 module  but also the module in the presence of different external stresses: DNA damage and oncogenic response. Second, using analytical arguments  we are able to clarify the role of the different parameters of the models in the stability of the fluctuations. Moreover, we show that this mechanism consistently reproduces some of the experimentally available data.

\subsection*{The p53-Mdm2 model and Mathematical Background} 

Probably the simplest version, and the starting point of any development of the p53-Mdm2 module is the loop presented in {\bf Figure 1}. This is a feedback loop in which the expression of p53 activates the presence of the Mdm2 that in turn regulates the activity of p53. This is a well studied model \cite{Zatorsky}, whose mean field solution is characterized by  damped oscillations. Sometimes, it is convenient to introduce an intermediary for the interaction between the p53 and the Mdm2, usually a messenger mRNA that acts as the transcription factor of the Mdm2. This intermediary is the responsible for a dephasing between the oscillations of the p53 and Mdm2 proteins \cite{RCF} but for simplicity we will assume in the rest of the work that it is in equilibrium, keeping, as our basic block for the p53-Mdm2 module, the scheme of {\bf Figure 1}.

We formalize the model starting from a stochastic approach  considering the number of components of each species, p53 ($n_{p53}$) and Mdm2 ($n_{Mdm2}$) as the relevant variables of the problem. 

From this point of view one faces the problem of solving the following Master equation:

\begin{equation}
\frac{\partial P(\vec{n},t)}{\partial t}= \sum_{\vec{n}\neq\vec{n'}} 
(P(\vec{n'},t)W_{\vec{n'}\rightarrow\vec{n}}-P(\vec{n},t)W_{\vec{n}\rightarrow\vec{n'}})
\label{eq:ecuacionmaestra}
\end{equation}

\noindent where $P(\vec{n},t)$ is the probability that the system is in state $\vec{n}=(n_{p53},n_{Mdm2})$ at time $t$ and $W_{\vec{n}\rightarrow\vec{n'}}$ are transition rates from one state to the other. The solution of eq. (\ref{eq:ecuacionmaestra}) is not an easy task, but, for very simple transition rates and one most usually use numerical methods, or resort to approximations. In this work we try both approaches and compare the results.

The numerical solution is obtained using the Gillespie's algorithm\cite{Gillespie}. As other Monte Carlo techniques this algorithm guarantees an exact solution, but it is time consuming and is not a good tool to explore the parameter space. To understand the role of the parameters of the model in the sustainability of the fluctuations, it is important to have at least approximate solutions to the problem. Here,
we use the Linear Noise Approximation (LNA) \cite{vanKampenlibro}, that was first proposed in the context of chemical kinetics and have gain considerable attention in the last few years for modeling intracellular processes, \cite{ElfEhren,Scott2}, but also in more general contexts \cite{McKane,Nuris}. The mathematical details of this approximation can be found in \cite{vanKampenlibro}, and we summarize them in section  {\bf Methods}. However, to simplify the reading, and to keep the consistency of the manuscript  we present here its main results.

 One of the difficulties to solve  equation (\ref{eq:ecuacionmaestra}) is the discrete character of $n_i$. To deal with this, van Kampen\cite{vanKampenlibro} proposed the following approximation:

\begin{equation}
n_i = \Omega x_i + \sqrt{\Omega} \alpha_i
\label{eq:ni0}
\end{equation}

\noindent i.e., one approximates the number of {\em molecules} of specie $i$ by some concentration, plus some fluctuations proportional to the square root of the system size $\Omega$.  Substituting (\ref{eq:ni0}) into (\ref{eq:ecuacionmaestra}) and developing the equation in order of $\frac{1}{\Omega}$ one obtains, at first order a set of differential equations for $x_i$, usually called the mean field or deterministic approximation:
\begin{equation}
\frac{\partial x_i}{\partial \tau} = \sum_j S_{ij} \hat{W_j}(\vec{x} \Omega)
\label{eq:mfield}
\end{equation}

\noindent where $S_{ij}$ is the stoichiometric matrix of the interactions (see the section {\bf Methods} for a detail derivation). At second order one can prove that the fluctuations $\alpha_i$ are Gaussian variables. However, to compare them with the experiments it is more useful to describe the evolution of these variables in  terms of Langevin equations.

\begin{equation}
\frac{\partial \alpha_{i}}{\partial \tau}=\sum_{k=1}^{N}A_{ik}\alpha_{k}+\eta_{i}(\tau)
\label{eq:langevin}
\end{equation}

\noindent where $A_{ik} = \sum_{j=1}^{M}S_{ik}\frac{\partial W _{ij}}{\partial x_{k}}$ and the $\eta_i$ has zero variance and correlation function $\langle \eta_i(\tau) \eta_j(\tau') \rangle = D_{ik} \delta(\tau-\tau)$ with $D_{ik}=\sum_{j=1}^{M}S_{ij}S_{kj}W_{j}(\vec{x})$  (see {\bf Methods}). 

Equation (\ref{eq:langevin}) is a linear Langevin equation that can be easily solved in the Fourier representation

\begin{equation}
\hat{\alpha}_i(w) = \sum_k \Phi_{ik}^{-1} \hat{\eta}_k(w)
\end{equation}

\noindent where $\Phi_{ik} = i w \delta_{ik} - A_{ik}$. From $\hat{\alpha}_i(w)$ one can get the power spectrum of the fluctuations as:

\begin{equation}
P_i(w) = \langle | \tilde{\alpha}_i(w) |^2 \rangle = \sum_j \sum_k \Phi_{ij}^{-1*}\Phi_{ik}^{-1} D_{jk}
\label{eq:power}
\end{equation}

\noindent which,  when studying fluctuations in single cells, is the quantity usually reported experimentally \cite{Zatorsky,Loewer}.


\section*{Results and Discussion}

In this section we present the main results of our work. It is divided in three parts, in each one of them the p53-Mdm2 module is studied either as an isolated structure or considering its interactions with different external elements. The organization of each subsection is the same, we first motivate the model from the biological point of view and then present its stochastic formulation.  We will show the results of simulations using Gillespie's algorithm and compare them with the prediction of the corresponding mean field solution (\ref{eq:mfield}). Then, we study the fluctuation spectra obtained from the simulations and compare them with the  predictions of the Linear Noise Approximation (LNA). Finally when it is possible we discuss the connection with the experimental results.

\subsection*{Basal response}

In the absence of any external signals the p53, is activated only by random DNA damage. This is a common event during the cell cycle life. Spontaneous hydrolysis, collapsing replication forks and oxidative stress \cite{Branzei,Sancar}, but also metabolic stress may produce p53 activation \cite{Karen}. Within this context we study the basal response using the model in {\bf Figure 1}. 

In stochastic terms the model is well described by the following set of expressions:

\begin{equation}
p53 \overset{W_{1}}{\longrightarrow}p53+Mdm2\label{reac1}
\end{equation}
\begin{equation}
p53\overset{W_{2}}{\longrightarrow}2p53\label{reac2}
\end{equation}
\begin{equation}
p53\overset{W_{3}}{\longrightarrow}\Phi\label{reac3}
\end{equation}
\begin{equation}
Mdm2\overset{W_{4}}{\longrightarrow}\Phi\label{reac4}
\end{equation}
\begin{equation}
p53+Mdm2\overset{W_{5}}{\longrightarrow}Mdm2\label{reac5}
\end{equation}

\noindent where equations (\ref{reac1}) and (\ref{reac2}) reflect the activation of Mdm2 and the self-activation of p53. Equations (\ref{reac3}) and (\ref{reac4}) the decay of both species and equation (\ref{reac5}) reflects the regulation of p53 by Mdm2. The specie $\Phi$ is included to allow the free variation of the quantities of p53 and Mdm2 during the simulation. The different $W_i$ are transition probabilities proportional to the size of the system \cite{vanKampenlibro}. To fix concepts we fixed the simplest possible transition probabilities, in this case:

\begin{eqnarray}
W_1 = k_1 \Omega x_{p53}\\
W_2 = k_2 \Omega x_{p53}\\
W_3 = k_3 \Omega x_{p53}\\
W_4 = k_4 \Omega x_{Mdm2}\\
W_5 = k_5 \Omega x_{p53} x_{Mdm2} 
\end{eqnarray}

 One should notice that the transitions ($W_{1,2}$), i.e, activation of Mdm2, and self-activation of p53  are proportional to the quantity of p53 in the system. The degradation of Mdm2, ($W_4$) is proportional to the amount of Mdm2. The p53 is degraded following two mechanisms, a normal degradation ($W_3$) proportional to the quantity of p53 in the cell, and a degradation activated by Mdm2 ($W_5$), that  is proportional to the quantities of p53 and of Mdm2.

Following (\ref{eq:mfield})  the  mean-field equations of this model read:

\begin{eqnarray}
\frac{d x_{p53}}{dt} = k_2  x_{p53} - k_3  x_{p53} - k_5  x_{p53} x_{Mdm2}\\
\frac{d x_{Mdm2}}{dt} = k_1  x_{p53} - k_4 x_{Mdm2}
\label{eq:mfbasal}
\end{eqnarray}

The fixed points of these equations are: $(0, 0)$ and 
$(\frac{k_{4}k_{2}}{k_{1}k_{5}},\frac{k_{2}}{k_{5}}) $ where to simplify notation we considered $ k_2' = k_2-k_3 $ and relabeled it as  $k_2 $. Then, the non trivial solution has biological sense only if the rate of activation of p53 is larger than its degradation rate ($k_2>0$). 

The stability of these fixed points is defined by the following Lyapunov exponents:

\begin{equation}
	  (0,0) \rightarrow \vec{\lambda}=
	  \left\{ 
	  \begin{array}{c}
	      k_2 \\ -k_4
	  \end{array} 
	  \right\} 
	  \label{estabilidad de la trivial//oncogenes}
	    \end{equation}
	\smallskip
and:
  \begin{equation}
    (x^*_{p53}, x^*_{Mdm2}) \rightarrow \vec{\lambda} = \left\{ 
    \begin{array}{c}
   \frac{1}{2}(-k_4+\sqrt{k^2_4-4 k_2 k_4} )\\ \frac{1}{2}(-k_4-\sqrt{k^2_4-4 k_2 k_4})
   \end{array} \right\} \label{eq:stabilitybasal2}
\end{equation}
      
These results guarantee that if the non-trivial solution exists ( $k_4 >0$ and $k_2>0$ ), then it is stable. Moreover, if $k_4<4 k_2$, i.e. if the activation of p53 is large enough, it  displays damped oscillations.  This is the region of parameters relevant from the biological point of view.
    
 In {\bf Figure 2} we show the comparison between the numerical solution of these equations  with the results of the Gillespie's algorithm for the same values of the parameters. Note that, while the deterministic solution is over-damped, the stochastic solution displays persistent oscillations around the fixed point. 

To characterize these sustained oscillations we computed the power spectrum of the results of the Gillespie's simulations over 200 realizations  and made their geometrical average. The resulting power spectrum of the oscillations of the p53 is presented with points in  {\bf Figure 3} where we also show the analytical results predicted by the LNA (see equation (\ref{eq:power}) and section {\bf Methods}). This plot compares very well with {\em Figure 6} in \cite{Loewer} where the authors studied the basal response of the system and presented  experimental results for the Fourier transform of the fluctuations. 

The existence of a maximum in the power spectrum of the fluctuation in {\bf Figure 3} is associated to the presence of a characteristic frequency for these fluctuations.  In an infinite system, these oscillations die out, but the stochasticity produced by the finite size of the system acts as a white noise, see equation (\ref{eq:langevin}), that interacts with the proper frequency of the system producing a resonance-like effect. This characteristic frequency is connected with the presence of an imaginary part in the eigenvalues of the system of differential equations. 

This is made evident by a simple inspection of the LNA expression for the power spectrum of the model (see {\bf Methods}). There, one can easily check that the denominator has a minimum, very close to the expected frequency of the damped oscillations estimated in the mean field model. Below, we show that this phenomenon repeats in the other models, although in a more sophisticated form.

\subsection*{Response in the presence of stress}

The ATM is a kinase protein that is currently involved in the detection of the damage in the DNA. When the ATM detects the damage it induces the phosphorilation (activation) of p53.  We model the presence of this kind of external stress incorporating to the  module of the basal state the new protein (ATM) and its interaction with the p53 \cite{Zatorsky,Melisa}.

 In {\bf Figure 4} we present a sketch of this model.  From the technical point of view, we have now a problem where to the dynamics governing the basal state we must add the following new transitions:

\begin{equation}
ATM\overset{W_{6}}{\longrightarrow}ATM + p53 \label{reac6a}
\end{equation}
\begin{equation}
ATM + p53 \overset{W_{7}}{\longrightarrow}p53\label{reac7a}
\end{equation}
 \begin{equation}
ATM\overset{W_{8}}{\longrightarrow}\Phi\label{reac8a}
\end{equation}
 \begin{equation}
ATM\overset{W_{9}}{\longrightarrow}2 ATM \Phi\label{reac9a}
\end{equation}

These transitions reflect the activation of p53 by the ATM, ($W_6$), the regulation of the ATM by the p53, ($W_7$), the natural degradation of the ATM, ($W_8$), and a self-reinforcement field for the ATM, ($W_9$) that mimics the persistence of the damage. As in the previous section we assumed that they take the following simple form:

\begin{eqnarray}
W_6 = k_6 \Omega x_{ATM}\\
W_7 = k_7 \Omega x_{p53} x_{ATM}\\
W_8 = k_8 \Omega x_{ATM}\\
W_9 = k_9 \Omega x_{ATM}
\label{eq:transstressa}
\end{eqnarray}

With these definitions of the transition rules, we may write the mean field equations of the model: 

\begin{eqnarray}
&\frac{d x_{p53}}{dt} = k_2  x_{p53} - k_3  x_{p53} -& \\\nonumber
&k_5  x_{p53} x_{Mdm2} + k_6 x_{ATM}&\\\nonumber
&\frac{d x_{Mdm2}}{dt} = k_1  x_{p53} - k_4 x_{Mdm2}&\\\nonumber
&\frac{d x_{ATM}}{dt} = k_9  x_{ATM} - k_8 x_{ATM} - k_7 x_{ATM} x_{p53}&\nonumber
\label{eq:mfestressa}
\end{eqnarray}

These equations are similar to the ones in  (\ref{eq:mfbasal}), but now the role of ATM on the activation of p53 results in the addition of a new term to the first equation. Moreover, ATM  is also a new dynamical variable whose evolution must be taken into account.

This system of equations has three different fixed points. They are,  $(0,0,0)$, $(\frac{k_2 k_4}{k_1 k_5},\frac{k_2}{k_4},0)$ and  $(\frac{k_9}{k_7},\frac{k_9 k_1}{k_7 k_4},\frac{k_9}{k_7 k_6} (\frac{k_9 k_1 k_5}{k_7 k_4}-k_2))$  where for simplicity we re-labelled $ k_2'=k_2-k_3 $, and $ k_9'=k_9-k_8 $, as $k_2 $ and $k_9$ respectively.

The Jacobian matrix necessary to study their stability takes the general form:

	$$
	J =  \left[ \begin{array}{ccc}
		k_2-k_5 x^*_{Mdm2} & -k_5 x^*_{p53} & k_7\\
		k_1 & -k_4 & 0\\
		-k_7 x^*_{ATM} & 0 & 0 
		\end{array} \right] 
	$$

\noindent that evaluated in the fixed point of interest  $(\frac{k_9}{k_7},\frac{k_9 k_1}{k_7 k_4},\frac{k_9}{k_7 k_6} (\frac{k_9 k_1 k_5}{k_7 k_4}-k_2))$
transforms into the following secular equation:
          \begin{equation}
	    \lambda^3+\lambda^2(c-k_4)+\lambda(\frac{k_9 k_1 k_5}{k_7}-k_4 c-k_9 c)-k_9 k_4 c=0 \label{eq:secular}
	  \end{equation}

\noindent where $c = k_2 - \frac{k_9 k_1 k_5}{k_7 k_4}$ . 

This secular equation must be studied numerically and its solution defines the stability of the fixed points. We have found that depending on the parameters, three different regions are well defined. In Region I the solution $\{x^*_{p53}, x^*_{Mdm2}, x^*_{ATM}\}$ may exist,  is stable and the system reaches the fixed point with damped oscillations. In this region also the solution  $\{x^*_{p53}, x^*_{Mdm2}, 0\}$ is stable if $k_{2}>0$. Depending on the initial condition, the system is attracted to one or another fixed point. 

In Region II the solution $\{x^*_{p53}, x^*_{Mdm2}, x^*_{ATM}\}$ has biological sense and is stable, but the fixed point is reached without damped oscillations. In this  region the solution $\{x^*_{p53}, x^*_{Mdm2}, 0\}$ does not exist. Finally, in Region III only the solution  $\{x^*_{p53}, x^*_{Mdm2}, 0\}$ is stable.

On the basis of the results obtained studying our previous model we concentrate our attention in Region I, where the mean field fixed point  $\{x^*_{p53}, x^*_{Mdm2}, x^*_{ATM}\}$ is reached through damped oscillations. In {\bf Figure 5} we show  the numerical solution of  equation (\ref{eq:mfestressa}) and the results of the  
Gillespie's algorithm using the parameters $\vec{k} =(0.99, 1, 0.44, 0.69, 0.85, 0.5, 0.5, 0.1, 0.4) h^{-1}$. As in the previous section, we see that the stochastic simulation shows sustained oscillations in time.

To characterize these oscillations we use again the power spectrum of the Gillespie's results. 
In {\bf Figure 6} we show with points this power spectrum and with a continuous line the predictions from the LNA. Note again the good coincidence, but one can also compare them with {\em Figure 3} in \cite{Zatorsky}. Our model reproduces qualitatively and quantitatively the experimental observations: the existence of a characteristic frequency  and a divergence close to zero frequency. 

The peak in the power spectrum still reflects the resonance-like mechanism between the p53-Mdm2 feed-back loop and the intrinsic noise. On the other hand, the divergency at low frequencies results from the presence of the new interaction and can be interpret looking to the analytical form of the power spectrum derived in the LNA at low frequencies. It has the form

\begin{equation}
P_{p53} (w \rightarrow 0) \sim \frac{2 k_5 k_7}{k_1 k_6 k_{13} - k_2 k_5 k_9}
\end{equation}

\noindent and guarantees large values of the power spectrum if the denominator is small enough.

\subsection*{Oncogenic response}

Another important pathway for the activation of p53, comes from the activation of the p14$^{ARF}$ protein. In the presence of oncogenes this protein is activated and joins the Mdm2 accelerating its degradation \cite{Sandra}. This process leads to the increase of the p53 level in the cell.

As far as we know there is no experimental evidence for the single cell dynamics of p53 and Mdm2 in the presence of oncogenes, therefore the parameters used here have no experimental bias. Nevertheless we will show that independently of the values used the power spectrum of the fluctuations is qualitatively different from the one obtained when the p53 interacts directly with ATM.  

The model of oncogenic response is sketched in {\bf Figure 7}. To the standard interactions present in the basal model we add the regulation of p14$^{ARF}$ to Mdm2 and a regulation of $p14^{ARF}$ by p53. In terms of stochastic transitions we have now:

\begin{equation}
Mdm2 + p14^{ARF}\overset{W_{6}}{\longrightarrow}p14^{ARF} \label{reac6b}
\end{equation}
\begin{equation}
p14^{ARF} + p53 \overset{W_{7}}{\longrightarrow}p53\label{reac7b}
\end{equation}
 \begin{equation}
p14^{ARF}\overset{W_{8}}{\longrightarrow}\Phi\label{reac8b}
\end{equation}
 \begin{equation}
p14^{ARF}\overset{W_{9}}{\longrightarrow}2 p14^{ARF} \Phi\label{reac9b}
\end{equation}

\noindent and defining the transition rates as:

\begin{eqnarray}
W_6 = k_6 \Omega x_{p14} x_{Mdm2}\\
W_7 = k_7 \Omega x_{p53} x_{p14}\\
W_8 = k_8 \Omega x_{p14}\\
W_9 = k_9 \Omega x_{p14}
\label{eq:transstressb}
\end{eqnarray}

\noindent we obtain the following system of equations for the dynamic of the system:

\begin{eqnarray}
&\frac{d x_{p53}}{dt} = k_2  x_{p53} - k_3  x_{p53} - k_5  x_{p53} x_{Mdm2}& \\\nonumber
&+ k_6 x_{ATM}&\\\nonumber
&\frac{d x_{Mdm2}}{dt} = k_1  x_{p53} - k_4 x_{Mdm2} - k_6  x_{p14} x_{Mdm2}&\\\nonumber
&\frac{d x_{p14}}{dt} = k_9  x_{p14} - k_8 x_{p14} - k_7 x_{p14} x_{p53}&\nonumber
\label{eq:mfestressb}
\end{eqnarray}

\noindent which again depending on the parameters may show three fixed points: $(0,0,0)$, $(\frac{k_2 k_4}{k_1 k_5},\frac{k_1}{k_5},0)$ and
$ (\frac{k_9}{k_7},\frac{k_2}{k_5},\frac{1}{k_2 k_6} (\frac{k_9 k_1 k_5}{k_7 }-k_2 k_4))$ where to simplify the notation we used $k_2=k_2-k_3$ and 
$k_9 = k_9-k_7$. 

The Jacobian matrix near these fixed points takes the form:

\begin{equation}
	J =  \left[ \begin{array}{ccc}
		k_2-k_5 x^*_{Mdm2} & -k_5 x^*_{p53} & 0\\
		k_1 & -k_4 & -k_6 x^*_{Mdm2}\\
		-k_2 x^*_{p14} & 0 & k_9 
		\end{array} \right] 
\end{equation}

 \noindent that evaluated in the fixed point  $\{x^*_{p53}, x^*_{Mdm2}, x^*_{p14}\}$ leads to the following secular equation:
 
\begin{equation}
	    \lambda^3+\lambda^2k_4+\lambda(\frac{k_9 k_1 k_5}{k_7})+\frac{k^2_9 k_5 k_1}{k_7}-k_2k_9 k_4 =0 
	    \label{ estabilidad de la 3}
 \end{equation}	

\noindent which defines a phase diagram qualitatively similar to the one discussed in the preceding section. Three regions, but only in one of them, the solution $\{x^*_{p53}, x^*_{Mdm2}, x^*_{p14}\}$ displays damped oscillations.
Our previous results suggest that this is the relevant region for our purposes. 

Then , using as parameters $\vec{k} = ( 1.0,1.0,0.44,0.69,0.85,0.5,0.5,0.4)$ we obtain damped oscillations in the mean field solution of the problem and sustained oscillations through the Gillespie's  algorithm. The results are shown in {\bf Figure 8}.

In {\bf Figure 9} we show the power spectrum obtained using the LNA and the one obtained from the simulations. Also for this model, the LNA approximation describes correctly the results of the simulation. As before a clear peak representative of the feed-back loop between the p53 and Mdm2 is observed, however, there is no evidence of a divergence at small frequencies. 

This absence of the divergence is independent of the parameters chosen. It  can be understood looking again to the structure of the power spectrum predicted by the LNA at small frequencies. For this model:

\begin{equation}
P_{p53} (w \rightarrow 0) = \frac{2 k_5 (k_1 + k_2)}{k_1^2 k_6}
\end{equation}

 \noindent that indicates that only when $k_1$ or $k_6$ are zero, i.e, only when the regulation of the Mdm2 by the p53 or the interaction between the Mdm2 and p14$^ARF$ disappears we have a diverging power spectrum at small frequencies. But in this case our model of interactions breaks down.

\section*{Conclusions}

We studied, using analytical tools (LNA) and numerical simulations (Gillespie's algorithm) the p53-Mdm2 module in its basal state, in the presence of DNA damage and oncogenic signals. The module is studied as a stochastic system where fluctuations are relevant.  Our  results compare quantitatively well with the experimental data available. This suggests that the intrinsic noise due to the finite size of the populations of p53 and Mdm2 may be responsible of the sustained oscillations in the response of the p53-Mdm2 circuit.

\section*{Methods}
\label{sec:Methods}

\subsection*{Linear Noise Approximation}
\label{App:LNA}

The quantities characterizing the number of particles in our problem are discrete. The deviation of this number from the expected mean value produces fluctuations usually called {\em intrinsic noise} because it is not caused by any external source. When the system is small enough, this intrinsic noise is not negligible, and a mean field approximation can not be taken.

In these cases, the correct way to describe the system is to solve the Master Equation that governs it:

\begin{equation}
\frac{\partial P(\vec{n},t)}{\partial t}= \sum_{\vec{n}\neq\vec{n'}} 
(P(\vec{n'},t)W_{\vec{n'}\rightarrow\vec{n}}-P(\vec{n},t)W_{\vec{n}\rightarrow\vec{n'}}) \mbox{.}
\label{eq:maestra}
\end{equation}

This is an equation for the change in time of the probability $P(\vec{n},t)$ of being in the state $\vec{n}$ at the time $t$, and it rests upon the Markovian hypothesis. $W_{\vec{n}\rightarrow\vec{n'}}$ are transition rates from states $\vec{n}$ to $\vec{n'}$.

An important difficulty in solving the Master equation arises from the discreteness of the variables involved ($\vec{n}$). Its solution is obtained by analytical methods in very lucky situations. An approximation scheme that allows to get some information from it, is the linear noise approximation (LNA). 

This approximation assumes the possibility of studying the number $n_i$ of integrants of the $i^{th}$ population as a main part $\Omega x_i$ that represents the mean value of the population size, plus a small deviation $\sqrt{\Omega} \alpha_i$ that represents the fluctuations around the mean value, and that is proportional to the square root of the system size $\Omega$. In this way:

\begin{equation}
n_i = \Omega x_i + \sqrt{\Omega} \alpha_i \mbox{.}
\label{eq:ni}
\end{equation}

As every process that takes place in the system causes variations of one or more integrands in the populations of the species involved, it is useful to write the master equation in terms of these discrete fluctuations. For this purpose let us define the {\em step operator} as:

\begin{equation}
 E_i^{k} f(\ldots,n_i,\ldots) =  f(\ldots,n_i + k,\ldots)\mbox{.}
\end{equation}
In terms of this operator the master equation reads:

\begin{equation}
 \frac{dP(\vec{n},t)}{dt} =  \sum_{j=1}^{M} \left[\left( \prod_{i=1}^{N}E_i^{-S_{ij}}\right)-1 \right] W_j(\vec{n},\Omega)P(\vec{n},t) \mbox{,}
\label{ecuacionmaestraE}
\end{equation}
where $W_j$ is the microscopic reaction rate for the $j^{th}$ process and $S_{ij}$ is the element $(i,j)$ of the stoichiometry matrix $\textbf{S}$, the magnitude of the change in the $i^{th}$ population when the $j^{th}$ process takes place. The total number of different populations that form the system is $N$, and $M$ is the total number of different kinds of processes that can take place in the system. 

In terms of the fluctuations, which are continuous variables, the step operator takes the following form, more suitable for an analytic treatment

\begin{equation}
 {E_i}^k \approx 1 + \frac{k}{\sqrt{\Omega}} \frac{\partial  }{\partial \alpha_i} + \frac{k^{2}}{2\Omega}\frac{\partial^{2}}{\partial {\alpha_i}^{2}} + \ldots
\end{equation}
in this way, we can work with a continuous operator, and still not lose the stochastic features of these processes. With it 

\begin{equation}
\begin{split}
    \prod_{i=1}^{N}E_i^{-S_{ij}} f_j(\vec{\alpha},\Omega) = \left[ 1-\Omega^{-\frac{1}{2}}\sum_{i=1}^{N} S_{ij} \frac{\partial}{\partial \alpha_i}\right]f_j(\vec{\alpha},\Omega) +\\
     \frac{1}{2} \left[ \Omega^{-1} \sum_{i=1}^{N} \sum_{k=1}^{N} S_{ij} S_{kj}\frac{\partial}{\partial \alpha_i}\frac{\partial}{\partial \alpha_k} + \ldots \right]  f_j(\vec{\alpha},\Omega)
     \end{split}
\label{productodeE}
\end{equation}

Now we are going to replace the probability $P(\vec{n},t)$ by the probability  $\Pi(\vec{\alpha},t)$ of being $\sqrt{\Omega}\vec{\alpha}$ away from the mean field prediction
\begin{equation}
 P(\vec{n},t)=P(\Omega \vec{x}+\sqrt{\Omega}\vec{\alpha})=\Pi(\vec{\alpha},t)
\label{PyPi}
\end{equation}

We know that:

\begin{eqnarray*}
  \frac{\partial\Pi(\vec{\alpha},t)}{\partial t} = \frac{\partial P(\vec{n},t)}{\partial t}+ \Omega \sum_{i=1}^{N} \frac{\partial P(\vec{n},t)}{\partial n_i} \frac{\partial x_i}{\partial t}  \;\;\;\;\;\;\ \mbox{and} \\
   \;\;\;\;\;\;\
 \frac{\partial P(\vec{n},t)}{\partial n_i} = \frac{1}{\sqrt{\Omega}} \frac{\partial \Pi(\vec{\alpha},t)}{\partial \alpha_i}
\mbox{,}
\end{eqnarray*}
with this results we can obtain the relation
\begin{equation}
\begin{split}
  \frac{\partial\Pi(\vec{\alpha},t)}{\partial t}-\Omega^{\frac{1}{2}}\sum_{i=1}^{N} \frac{\partial \Pi(\vec{\alpha},t)}{\partial \alpha_i}  \frac{\partial x_i}{\partial t} = \frac{\partial P(\vec{n},t)}{\partial t}
\end{split}
\label{derivadasdePyPi} \mbox{.}
\end{equation}

If now we expand the transition probabilities per unit time around $\vec{x}$

\begin{equation}
\begin{split}
 W_j(\vec{\alpha},\Omega)= W_j(\vec{x}+\Omega^{-\frac{1}{2}} \vec{\alpha}) \simeq W_j(\vec{x}) + \\
 \Omega^{-\frac{1}{2}} \sum_{i=1}^{N} \frac{\partial W_j(\vec{x})}{\partial x_i} \alpha_i + o(\Omega^{-1})
\end{split}
\label{desarrollodeW}
\end{equation}

Substituting (\ref{productodeE}) and (\ref{desarrollodeW}) in the master equation (\ref{ecuacionmaestraE}) and using (\ref{PyPi}), after grouping the coefficients of the similar powers of $\Omega$ we find: 

\begin{equation}
\begin{split}
   \frac{\partial P(\vec{n},t)}{\partial t} = \Omega^{-1} \left( \sum_{j=1}^M W_j(\vec{x}) \sum_{i,k=1}^{N}  S_{ij} S_{kj} \frac{\partial^2 \Pi(\vec{\alpha},t)}{\partial \alpha_i \partial \alpha_k}\right) \\- 
 \Omega^{-1}\left(  \sum_{j=1}^M  \sum_{i,k=1}^{N} S_{ij} \frac{\partial[\Pi(\vec{\alpha},t) \alpha_k]}{\partial \alpha_i} \frac{\partial W_j(\vec{x})}{\partial x_k}\right)  \nonumber  \\ + \Omega^{-\frac{1}{2}}  \left( -\sum_{j=1}^M  \sum_{i=1}^{N} S_{ij} W_j(\vec{x}) \frac{\partial \Pi(\vec{\alpha},t) }{\partial \alpha_i} \right) + o(\Omega^{-\frac{3}{2}} )
\end{split}
\label{potenciasnonegativas}
 \end{equation}

If now we compare the coefficients of the similar powers of $\Omega$ in equations (\ref{potenciasnonegativas}) and (\ref{derivadasdePyPi}), where the unit of time has been changed from $t$ to $\tau= \frac{t}{\Omega}$ we find for the leading order:

\begin{equation}
 \frac{\partial x_i}{\partial \tau} = \sum_{j=1}^M S_{ij} W_j(\vec{x})\mbox{.}
\end{equation}

This corresponds to the deterministic rate equations that are often used to describe the dynamics of such systems at a mean-field level. It has been discussed already in the subsection referring to mean field equations and stability analysis.

For the coefficients of $\Omega^{-1}$ the relation reads:

 \begin{equation}
 \begin{split}
 \frac{\partial \Pi(\alpha,\tau) }{\partial \tau} = -\sum_{i,k=1}^{N} A_{ik} \frac{\partial[\Pi(\alpha,t) \alpha_k]}{\partial \alpha_i} + \\
 \frac{1}{2} \sum_{i,k=1}^{N} D_{ik} \frac{\partial^2 \Pi(\alpha,t)}{\partial \alpha_i \partial \alpha_k} \mbox{,}
\end{split}
\label{FokkerPlank}
\end{equation}

where 
\begin{equation}
 A_{ik}  =  \sum_{j=1}^{M} S_{ij} \frac{\partial W_j(\vec{x})}{\partial x_k}
\end{equation}
is the jacobian matrix of the system, and 
\begin{equation}
 D_{ik}  =  \sum_{j=1}^{M} S_{ij} S_{kj} W_j(\vec{x}) \mbox{.}
\end{equation}

Equation (\ref{FokkerPlank}) is a Fokker-Plank equation for the probability for the system to have the deviation $\sqrt{\Omega}\vec{\alpha}$ around the mean-field prediction. This equation can be written  in a completely equivalent formulation more benevolent to investigation using Fourier transforms. The problem can then be formulated as the set of stochastic differential equations of the Langevin type:

\begin{equation}
 \frac{d \alpha_i}{d \tau} = \sum_{k=1}^N A_{ik} \alpha_k  + \eta_i(t)
\label{Langevin}
\end{equation}
where $\eta_i(\tau)$ is a Gaussian noise with $0$ mean and correlation given by $\left\langle \eta_i(\tau) \eta_k(\tau') \right\rangle = D_{ik} \delta (\tau-\tau')$.

This a linear Langevin equation that can be easily solved in the Fourier representation

\begin{equation}
\hat{\alpha}_i(w) = \sum_k \Phi_{ik}^{-1} \hat{\eta}_k(w)
\end{equation}

\noindent where $\Phi_{ik} = i w \delta_{ik} - A_{ik}$. From $\hat{\alpha}_i(w)$ one can get the the power spectrum of the fluctuations as:

\begin{equation}
P_i(w) = \langle | \tilde{\alpha}_i(w) |^2 \rangle = \sum_j \sum_k \Phi_{ij}^{-1*}\Phi_{ik}^{-1} D_{jk}
\label{eq:power2}
\end{equation}

\subsection*{Power spectrum in the basal state}
\label{App:PS}
Now we are going to analyze the general expression for the power spectrum (\ref{eq:power2}) for the basal state. The analytics for the other two models develops following the same steps, with a more cumbersome Algebra

\begin{equation}
P_{i}(\omega)\propto\overset{N}{\underset{j=1}{\sum}}\overset{N}{\underset{k=1}{\sum}}\Phi_{ij}^{-1}(\omega)D_{jk}(\Phi^{\dagger})_{ki}^{-1}(\omega)\label{espectrodepotenciafinal}\end{equation}

Since $\Phi=i\omega\delta_{ik}-A_{ik}$ and since $A$ and $D$ are
independent of $\omega$, $P_{i}(\omega)$ is a fraction, the numerator
and the denominator have both a polynomial structure of order $2N$. 
The explicit form of the denominator is: $\left|det\Phi(\omega)\right|^{2}$.

Then, we expect that the enhancement of the fluctuations would
be in the values of $\omega$ that minimize the denominator. Now we
are going to analyze the general expression for the denominator of
the power spectrum, and show explicit expressions for the basal response.

The matrix $A$ in the expression of $\Phi$ is the jacobian matrix
which can be written in terms of the eingenvalues, then we can write:

\begin{equation}
\begin{split}
det\Phi(\omega)=det\left(\begin{array}{cc}
i\omega-\lambda_{1} & 0\\
0 & i\omega-\lambda_{2}\end{array}\right)=\\
-\omega^{2}-i\omega(\lambda_{1}+\lambda_{2})+\lambda_{1}\lambda_{2}
\end{split}
\end{equation}

\begin{equation}
\begin{split}
P_{i}(\omega)\varpropto\frac{1}{(-\omega^{2}-i\omega(\lambda_{1}+\lambda_{2})+\lambda_{1}\lambda_{2})}\\
\frac{1}{(-\omega^{2}+i\omega(\lambda_{1}+\lambda_{2})+\lambda_{1}\lambda_{2})}
\end{split}
\end{equation}

\begin{equation}
\begin{split}
P_{i}(\omega)\varpropto\frac{1}{(\omega^{2}-\lambda_{1}\lambda_{2})^{2}+\omega^{2}(\lambda_{1}+\lambda_{2})^{2}}\\
=\frac{1}{(\omega^{2}-det(A))^{2}+\omega^{2}(Tr(A))^{2}}
\end{split}
\label{eq:expresiondep)}
\end{equation}

This form for the power spectrum shows clearly the existence of a
resonance: for a specific value of $\omega^{2}$ the denominator becomes small,
and the power spectrum has a large peak centered on this frequency.
The condition $dP_{i}(\omega)/d\omega=0$ gives:\begin{equation}
2(\omega^{2}-det(A))2\omega+2\omega(Tr(A))^{2}=0\label{eq:extremar}\end{equation}

Then the condition (\ref{eq:extremar}) simply becomes: $\omega^{2}=det(A)-\frac{(Tr(A))^{2}}{2}$,
it implies that: $2det(A)>(Tr(A))^{2}$, this condition in terms of
the stability matrix implies that the eigenvalues of $A$ are complex.
It means that if the power spectrum has an extreme the stability
matrix has complex eigenvalues.

Replacing in the expression for the power spectrum  the eigenvalues for the
model of the system in basal conditions: $\lambda_{1}=\frac{-k_{2}}{2}+i\sqrt{k_{2}k_{4}-\frac{k_{4}^{2}}{4}}$
and $\lambda_{2}=\frac{-k_{2}}{2}-i\sqrt{k_{4}k_{2}-\frac{k_{4}^{2}}{4}}$ 

\begin{equation}
P_{i}(\omega)\varpropto\frac{1}{\omega^{4}+\omega^{2}(k_{4}^{2}-2k_{4}k_{2})+k_{4}^{2}k_{2}^{2}}\label{eq:expresion de p(w)}\end{equation}

\begin{equation}
P_{i}^{'}(\omega)\varpropto\frac{4\omega^{3}+2\omega(k_{4}^{2}-2k_{4}k_{2})}{(\omega^{4}+\omega^{2}(k_{4}^{2}-2k_{4}k_{2})+k_{4}^{2}k_{2}^{2})^{2}}=0\end{equation}

$P_{i}(\omega)$ has an extreme in $\omega=\sqrt{k_{4}k_{2}-\frac{k_{4}^{2}}{2}}$

As we can see, the imaginary part of the eigenvalues $(Im[\lambda_{1,2}])$
and the value of frequency where $P_{i}(\omega)$ has an extreme
are similar. Now we are going to write it in a more clearly way:

\begin{equation}
\begin{split}
Im[\lambda_{1,2}]=\sqrt{k_{2}k_{4}-\frac{k_{4}^{2}}{4}}=\sqrt{k_{2}k_{4}}\sqrt{1-\frac{k_{4}^{2}}{4k_{2}k_{4}}}\\
\thickapprox\sqrt{k_{2}k_{4}}(1-\frac{k_{4}^{2}}{8k_{2}k_{4}})
\end{split}
\end{equation}

The above Taylor expansion may be a good approximation if we take into
account that for the corresponding values of the parameters $\frac{k_{4}^{2}}{4k_{2}k_{4}}<\frac{k_{4}^{2}}{2k_{2}k_{4}}<0.5$.

In the same way:
\begin{equation}
\begin{split}
\omega=\sqrt{k_{4}k_{2}-\frac{k_{4}^{2}}{2}}=\sqrt{k_{2}k_{4}}\sqrt{1-\frac{k_{4}^{2}}{2k_{2}k_{4}}}\\
\thickapprox\sqrt{k_{2}k_{4}}(1-\frac{k_{4}^{2}}{4k_{2}k_{4}})
\end{split}
\end{equation}

Summarizing, the power spectrum of the fluctuations has a
maximum at a frequency close to the imaginary part of the
eigenvalue that characterizes the frequency of the damped oscillations
in the deterministic regime.

\section*{Authors contributions}
L. C. and N. F. developed the program for running Gillespie's algorithm.  L.C, N. F and R. M. contributed to the construction of the models. L.C made the analysis of the mean field solutions and optimized the parameters to fit the experimental results. R. M. suggested the problem and drafted the manuscript. The three authors read and approved the final manuscript.

\section*{Acknowledgements}
 \ifthenelse{\boolean{publ}}{\small}{} 
RM was supported by the Alexander von Houmboldt Foundation at the University of Freiburg during the last stage of the manuscript. The authors want to acknowledge useful discussions with J. Pi\~nero during the early stage of this work.
 

{\ifthenelse{\boolean{publ}}{\footnotesize}{\small}
 \bibliographystyle{bmc_article}  
 \bibliography{bmc_p53article} }     


\ifthenelse{\boolean{publ}}{\end{multicols}}{}



\section*{Figures}

\subsection*{Figure 1 - p53-Mdm2 basic module}
\includegraphics[width=15cm]{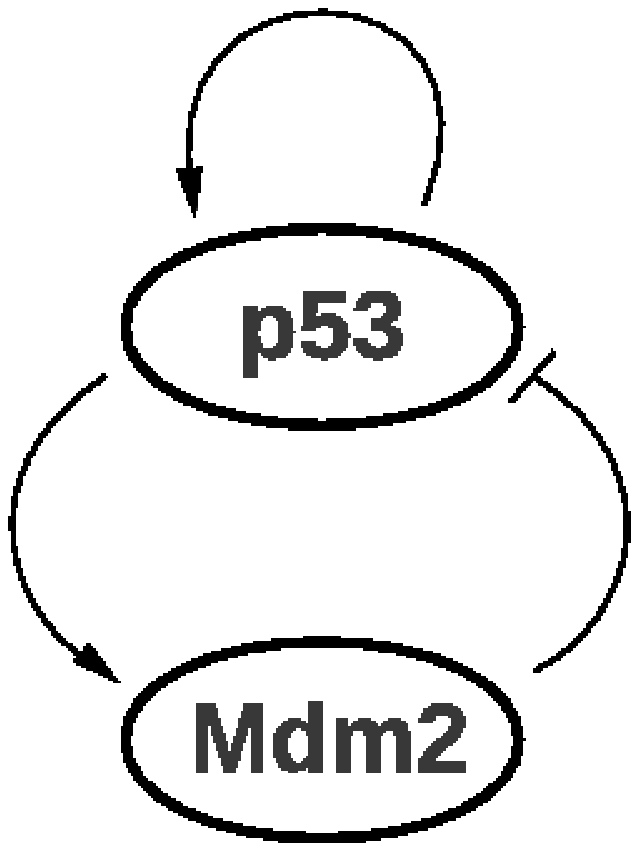}

Sketch of the p53-Mdm2 feedback loop in its basal state. The protein p53 activates Mdm2 and Mdm2 suppresses p53.

\subsection*{Figure 2 - Oscillations of the p53 in the basal state}
\includegraphics[width=15cm]{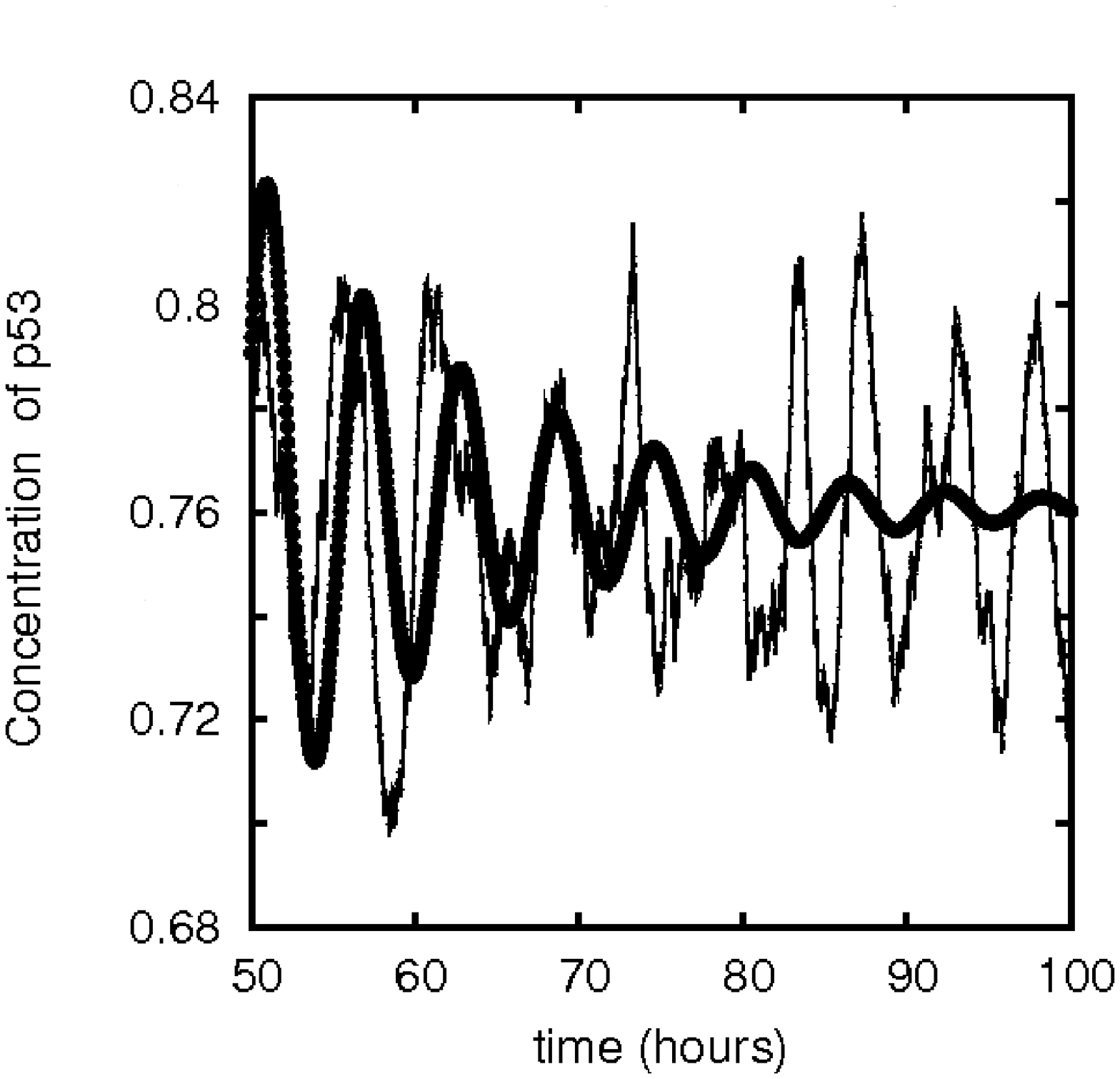}

Typical run showing the oscillations of the p53 in the basal state. The bold curve indicates the mean field solution and the thin one are results from the Gillespie's simulation.
Parameters: $\vec{k} =(0.99, 1, 0.44, 0.49, 1.05) h^{-1}$.

\subsection*{Figure 3 - Power spectrum of the fluctuations of the p53 in the basal state}
\includegraphics[width=15cm]{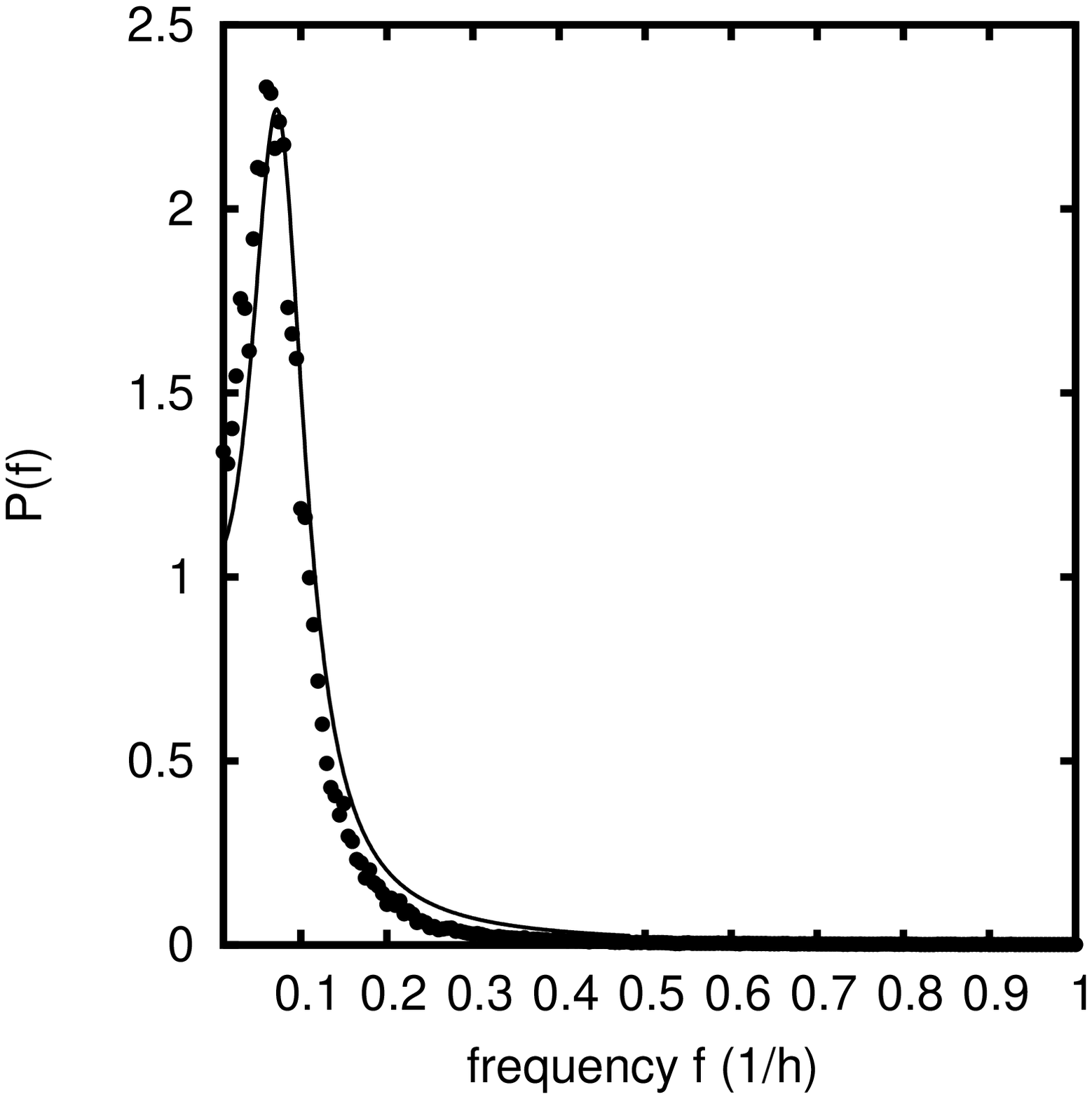}

Average power spectrum of the fluctuations of p53 in its basal state. The points represent the power spectrum obtained after averaging over 1000 realizations of the Gillespie's algorithm. The continue line is the prediction from the LNA. Note the existence of a characteristic frequency of oscillations.  
Parameters: $\vec{k} =(0.99, 1, 0.44, 0.49, 1.05) h^{-1}$.

\subsection*{Figure 4 - p53-Mdm2 model in the presence of DNA damage}
\includegraphics{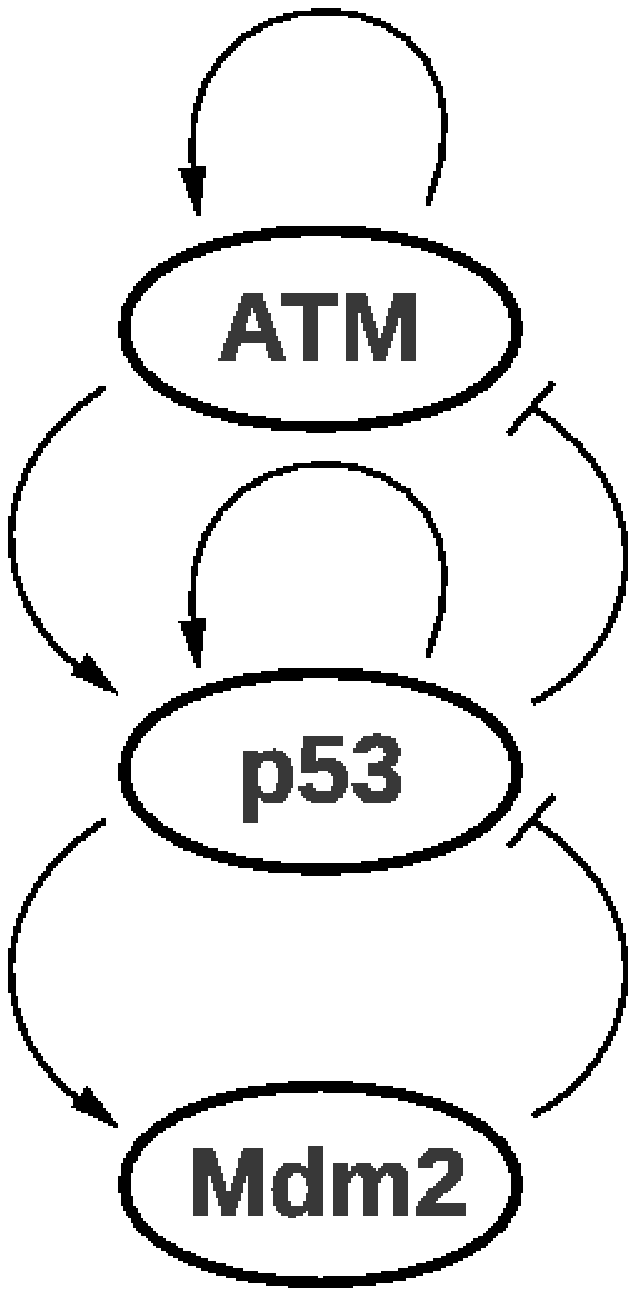}

Sketch of the p53-Mdm2 feedback loop in the presence of DNA damage. The protein p53 activates Mdm2 and Mdm2 suppresses p53. Morover. the ATM activates the p53 that in turn regulates ATM.

\subsection*{Figure 5 - Oscillations of the p53 in the  presence of DNA damage}
\includegraphics[width=15cm]{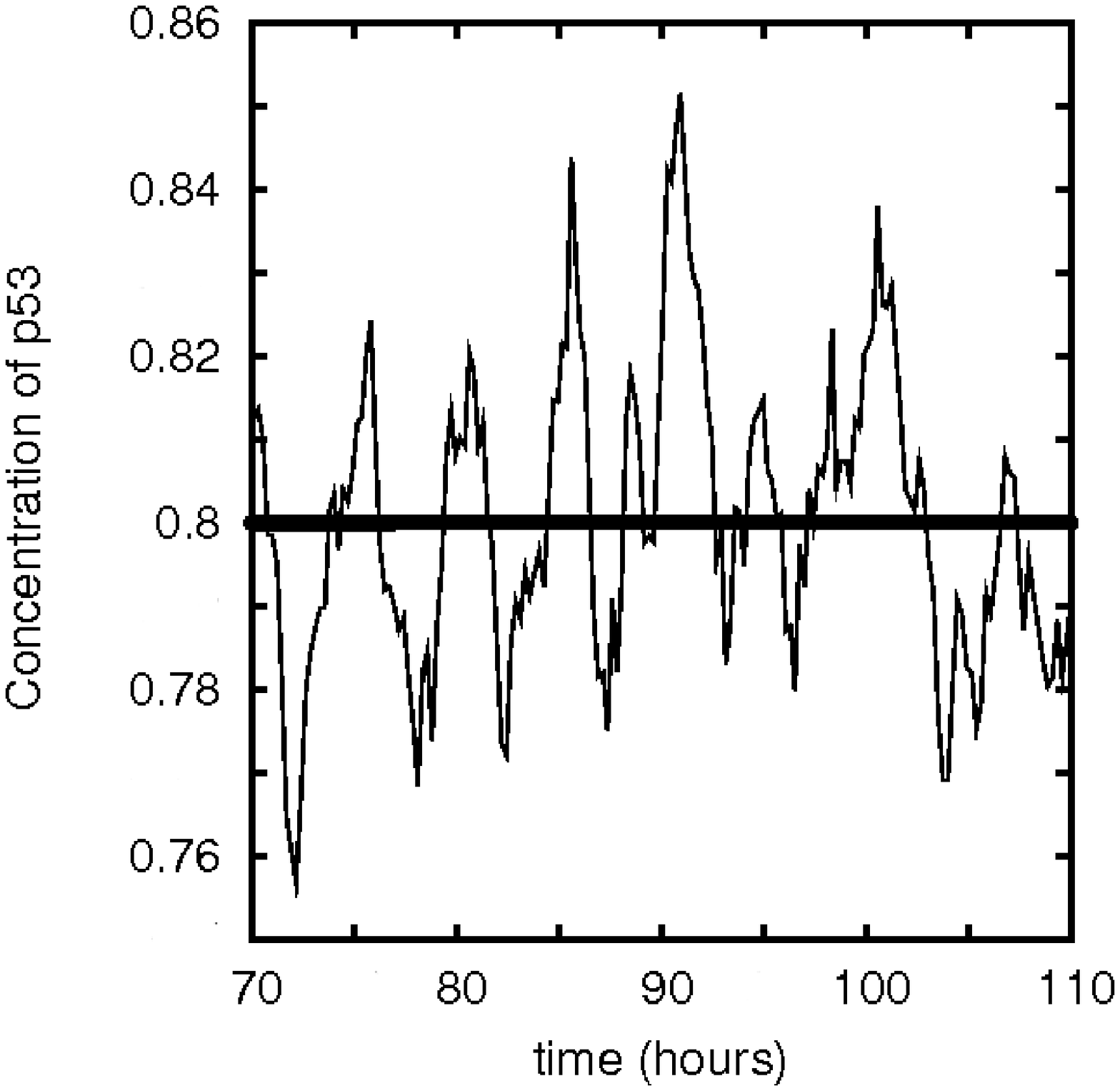}

Typical run showing the oscillations of the p53 in the presence of DNA damage. The bold curve indicates the mean field solution and the thin one are results from the Gillespie's simulation. 
Parameters: $\vec{k} =(0.99, 1, 0.44, 0.69, 0.85, 1.5, 1.5, 0.1, 1.0) h^{-1}$.

\subsection*{Figure 6 - Power spectrum  of the fluctuations of the p53 in the presence of DNA damage}
\includegraphics[width=15cm]{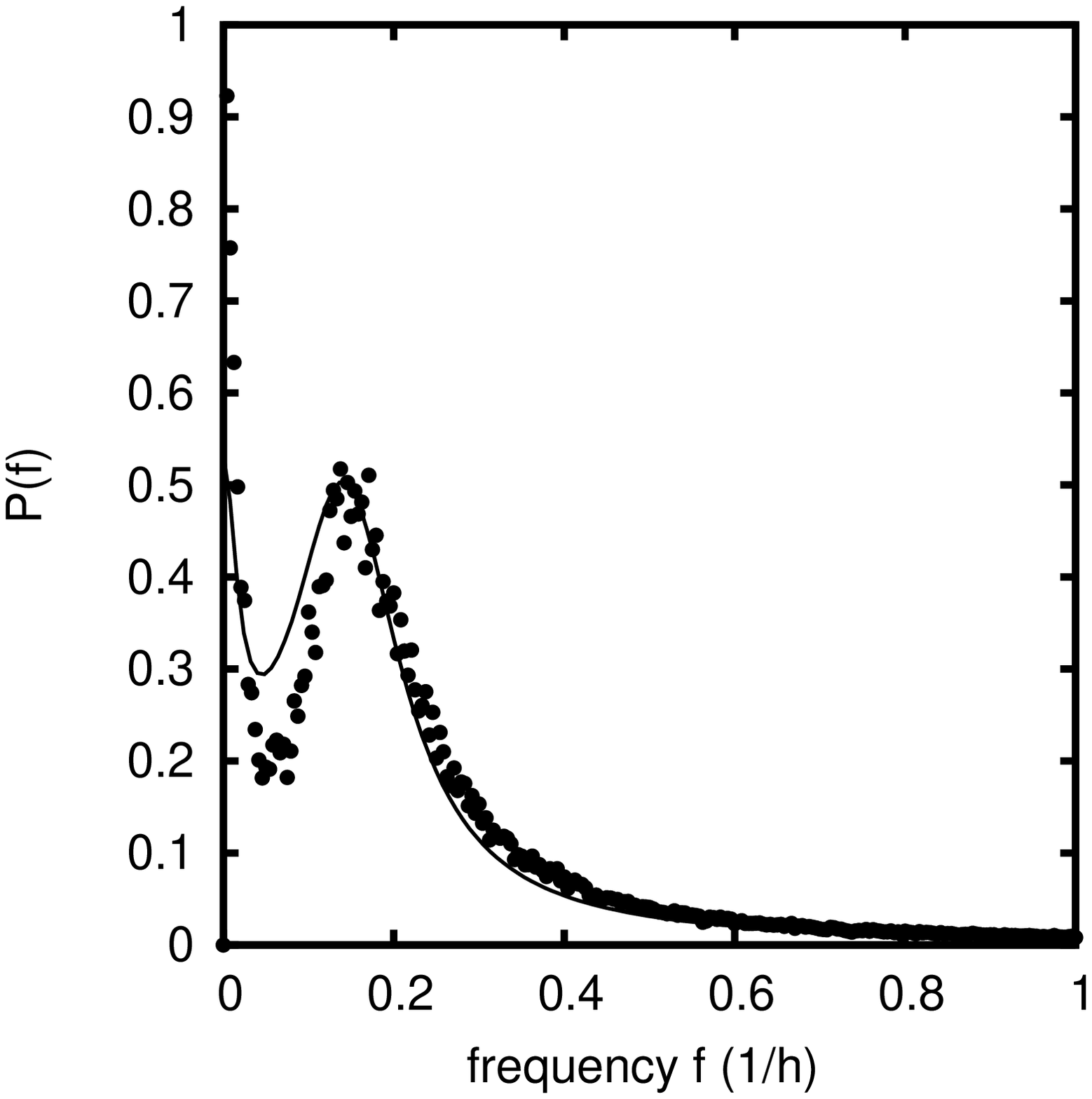}

The points represent the power spectrum obtained after averaging over 200 realizations of the Gillespie's algorithm. The continue line is the prediction from the LNA. Note the existence of a characteristic frequency of oscillations and a divergence close to zero frequency. 
Parameters:  $\vec{k} =(0.99, 1, 0.44, 0.69, 0.85, 1.5, 1.5, 0.1,1.0) h^{-1}$.

\subsection*{Figure 7 - p53-Mdm2 model for oncogenic response}
\includegraphics{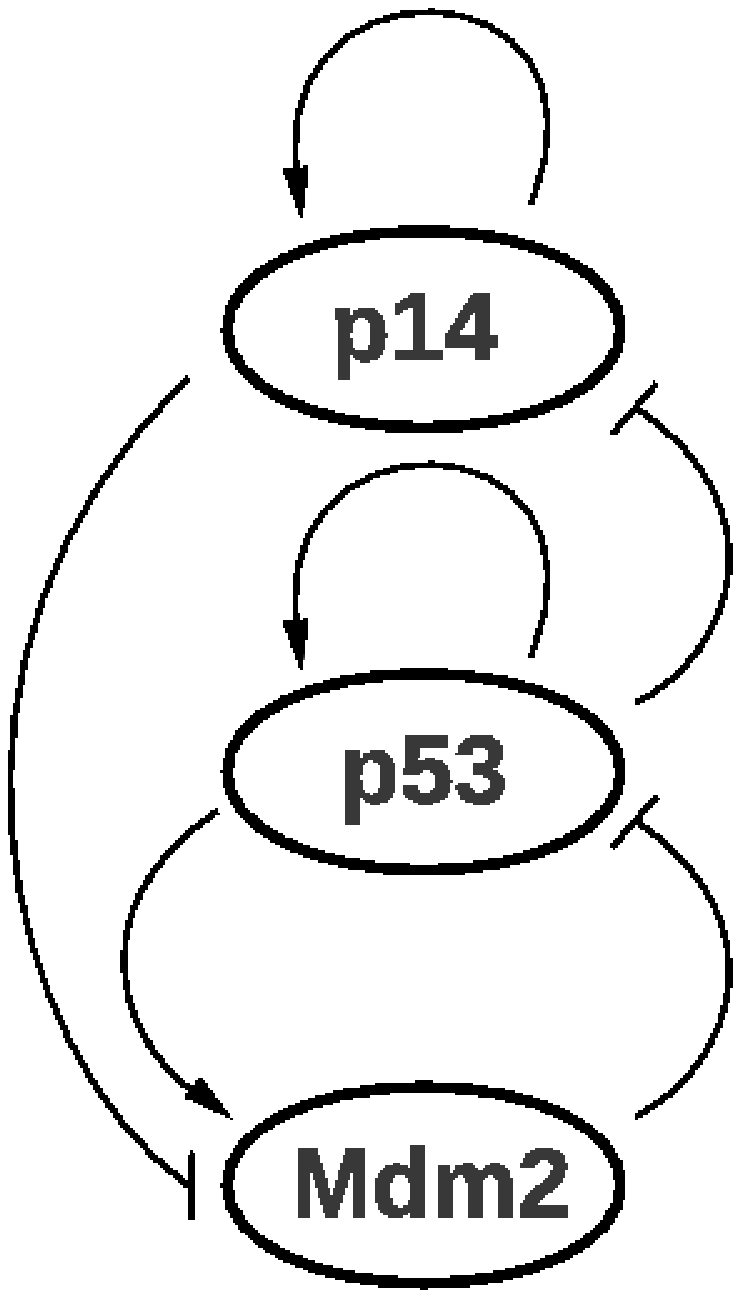} 

Sketch of the p53-Mdm2 feedback loop to model the oncogenic response. The protein p14 suppresses the Mdm2 and p53 suppresses p14.

\subsection*{Figure 8 - Oscillations  of the p53 after oncogenic response}
\includegraphics[width=15cm]{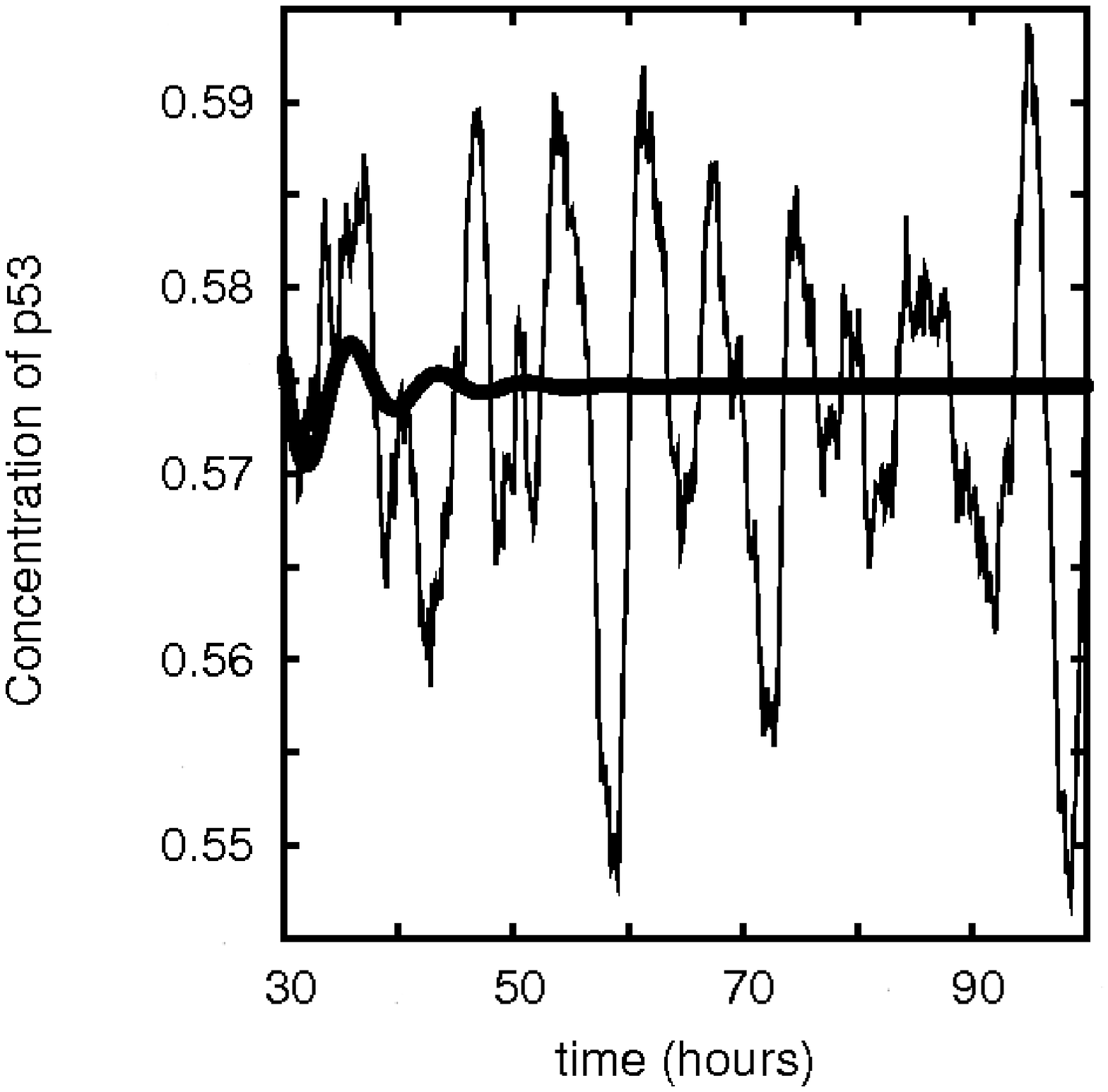}

Typical run showing the oscillations of the p53 in the presence of oncogenic signals. The bold curve indicates the mean field solution and the thin one are results from the Gillespie's simulation. 
Parameters: $\vec{k} = ( 1.0,1.0,0.44,0.69,0.85,0.5,0.5,0.4)$

\subsection*{Figure 9 - Power spectrum of the fluctuations of the p53 after oncogenic response}
\includegraphics[width=15cm]{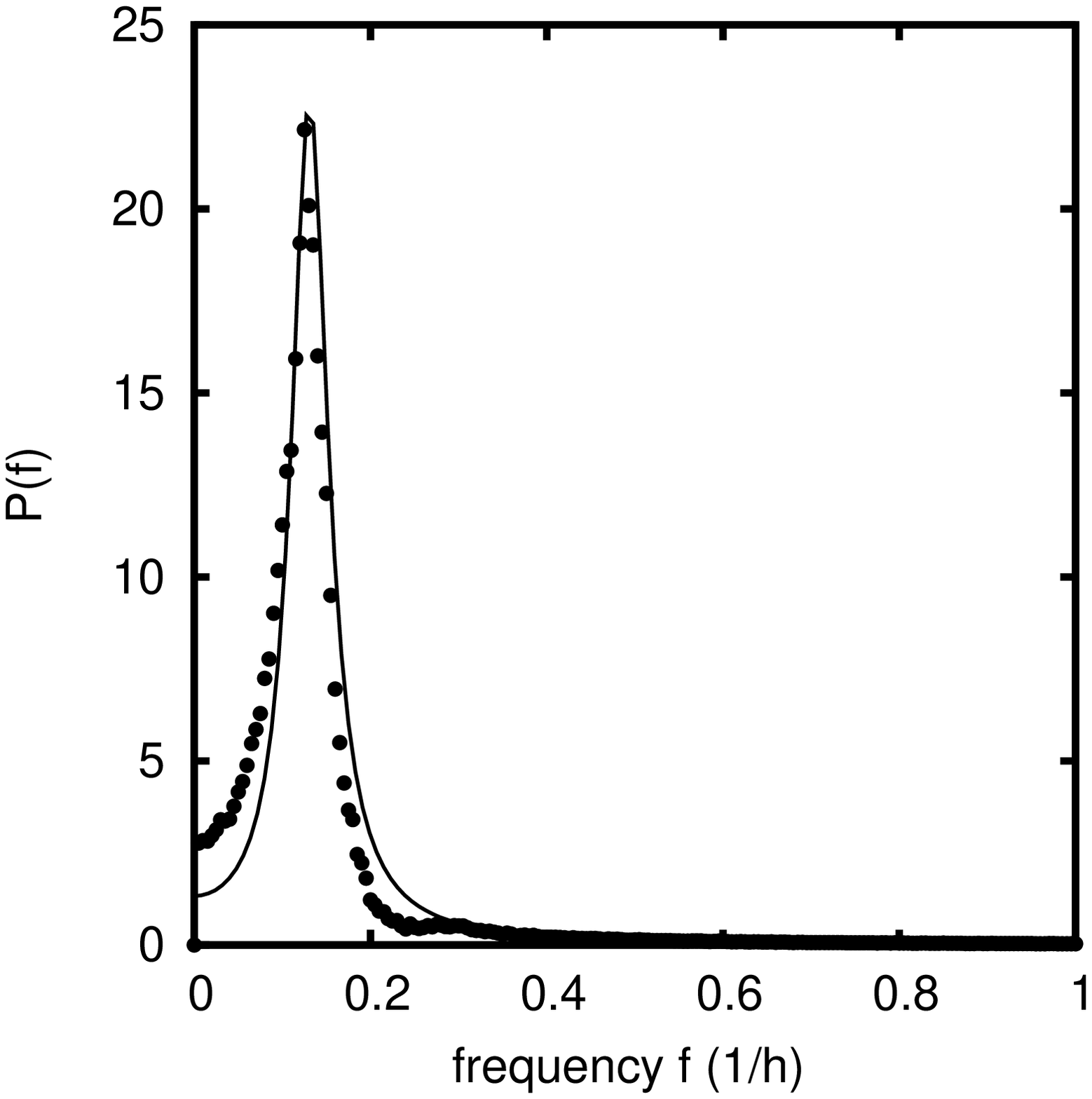}

The points represent the power spectrum obtained after averaging over 200 realizations of the Gillespie's algorithm. The continue line is the prediction from the LNA. Note the existence of a characteristic frequency of oscillations and a divergency close to zero frequency.
 Parameters: $\vec{k} = ( 1.0,1.0,0.44,0.69,0.85,0.5,0.5,0.4)$

\end{bmcformat}
\end{document}